\documentclass[useAMS]{emulateapj}
\usepackage[]{natbib}

\def\lsim{\lower.5ex\hbox{$\; \buildrel < \over \sim \;$}}
\def\gsim{\lower.5ex\hbox{$\; \buildrel > \over \sim \;$}}

\shorttitle{X-ray spectroscopy of the high mass X-ray binary Cen X-3}
\shortauthors{S. Naik, B. Paul and Z. Ali}

\begin{document}
\title{X-ray spectroscopy of the High Mass X-ray Binary Pulsar Centaurus~X-3 
over its binary orbit}   

\author{Sachindra Naik\altaffilmark{1}, Biswajit Paul\altaffilmark{2}, Zulfikar Ali\altaffilmark{1}}

\altaffiltext{1}{Astronomy \& Astrophysics Division, Physical Research Laboratory, Ahmedabad - 380009, India. $snaik@prl.res.in$}
\altaffiltext{2}{Raman Research Institute, Sadashivnagar, C. V. Raman Avenue, Bangalore 560080, India}

\begin{abstract}
We present a comprehensive spectral analysis of the high mass 
X-ray binary (HMXB) pulsar Centaurus~X-3 with the $Suzaku$ observatory
covering nearly one orbital period. The light curve shows the presence 
of extended dips which are rarely seen in HMXBs. These dips are seen 
up to as high as $\sim$40 keV. The pulsar spectra during the eclipse, 
out-of-eclipse, and dips are found to be well described by a partial 
covering power-law model with high energy cut-off and three Gaussian 
functions for 6.4 keV, 6.7 keV, and 6.97 keV iron emission lines. The 
dips in the light curve can be explained by the presence of an additional 
absorption component with high column density and covering fraction, the 
values of which are not significant during the rest of the orbital phases. 
The iron line parameters during the dips and eclipse are significantly 
different compared to those during the rest of the observation. During 
the dips, the iron line intensities are found to be lesser by a factor 
of 2--3 with significant increase in the line equivalent widths. However, 
the continuum flux at the corresponding orbital phase is estimated to be 
lesser by more than an order of magnitude. Similarities in the changes 
in the iron line flux and equivalent widths during the dips and eclipse 
segments suggests the dipping activity in Cen~X-3 is caused by obscuration 
of the neutron star by dense matter, probably structures in the outer 
region of the accretion disk, as in case of dipping low mass X-ray binaries. 
\end{abstract}


\section{Introduction}

Orbital phase dependent dipping activity in the X-ray binaries is believed 
to be caused by the obscuration of X-rays from the compact object by structures 
located in the outer regions of the disk particularly believed to be impact 
region of the accretion flow from the binary companion (White \& Swank 1982). 
Dipping in general leads to the hardening of the source spectrum because 
of the removal of the soft X-ray photons by photoelectric 
absorption. The observed properties during the dips viz. duration of the 
dip, depth of the dip, spectral behavior of the object during the dip 
etc. show strong variation from dip to dip. The dipping behavior repeats 
with the orbital period of the binary system. In the dipping low mass X-ray 
binary (LMXB) systems, the hardening of the energy spectrum during the dipping 
activities is often described by the ''progressive covering'' or ''complex 
continuum'' approach (Church et al. 1997). This approach explains the spectral 
changes during the dipping intervals by the partial and progressive covering by 
an opaque neutral absorber. It is found that the dipping activities in the X-ray 
intensity are very common in LMXBs (Parmar et al. 1999; Narita et al. 2003; Balman 
2009 and references therein). However, these activities are rarely seen in the 
high mass X-ray binary (HMXB) systems. 

In case of GX~1+4, though not an HMXB, intensity dips were seen in the X-ray 
light curves during which pulsation in soft X-rays were absent (Naik et al. 
2005 and references there in). These dips were interpreted as due to the 
absorption of soft X-ray photons by dense medium. In case of HMXB pulsar 
GX~301-2, an unusual intensity dip was observed in 1984 April. A drop in the 
X-ray continuum intensity (in 7--20 keV band) by an order of magnitude was 
detected which lasted for about 5 hrs (Leahy et al. 1988). The 6.4 keV iron 
fluorescent line flux was also reduced to about half with an increase of the 
equivalent width during the unusual event. The intensity dip observed in the 
non-eclipsing HMXB pulsar GX~301-2 was explained by an obscuration of the 
neutron star by dense matter which is smaller in size than the scattering 
region surrounding the neutron star. Dipping activity has been also recognized 
in 4U~1907+09 during which the X-ray intensity faded below the detection threshold
of the RXTE/PCA. These dips in 4U~1907+09 are interpreted as due to the cessations 
of the mass accretion by the neutron star (in't Zand et al. 1997). The dips
in Vela~X-1 are described as the onset of the propeller effect, which inhibits 
further mass accretion from the binary companion (Kreykenbohm et al. 2008).
This are the only known HMXB pulsars showing the presence of intensity dip
in the X-ray light curve.  Here is this paper, we discuss the presence of 
several intensity dips in the X-ray light curve of another HMXB pulsar
Cen~X-3 during its orbit using data from $Suzaku$ observatory. 

Cen~X-3 was the first binary pulsar to be discovered in X-rays (Giacconi 
et al. 1971). It is an eclipsing HMXB pulsar with a pulse period of $\sim$4.8 s 
and an orbital period of $\sim$2.1 days ((Schreier et al. 1972).  The 
binary system consists of a neutron star with a mass of 1.21$\pm$0.21 M$_\odot$ 
accompanied by an O 6-8 III supergiant star (V779~Cen) with a mass and radius 
of 20.5$\pm$0.7 M$_\odot$ and 12 R$_\odot$, respectively (Hutchings et al. 1979; 
Rappaport \& Joss 1983; Ash et al. 1999). The distance to the binary system was earlier 
estimated to be $\sim$8 kpc (Krzeminski 1974). However, using the energy-resolved 
dust-scattered X-ray halo, Thompson \& Rothschild (2009) estimated the distance
to Cen~X-3 to be 5.7$\pm$1.5 kpc. The high luminosity of the X-ray source ($\sim$5.0 
$\times$ 10$^{37}$ erg s$^{−1}$; Suchy et al. 2008) suggests that the predominant 
mode of accretion is via a disc, fed by incipient Roche-lobe overflow, although 
a strong stellar wind does emanate from the supergiant (Nagase et al. 1992; Day \&
Stevens 1993). The optical light curve (Tjemkes et al. 1986) supports the presence 
of an accretion disk, fed by Roche lobe overﬂow. Furthermore, Quasi Periodic 
Oscillations (QPOs) observed at $\sim$40 mHz (Takeshima et al. 1991; Raichur \& 
Paul 2008a) strengthen the case for the presence of an accretion disk.

\begin{figure*}
\centering
\includegraphics[height=5.6in, width=3in, angle=-90]{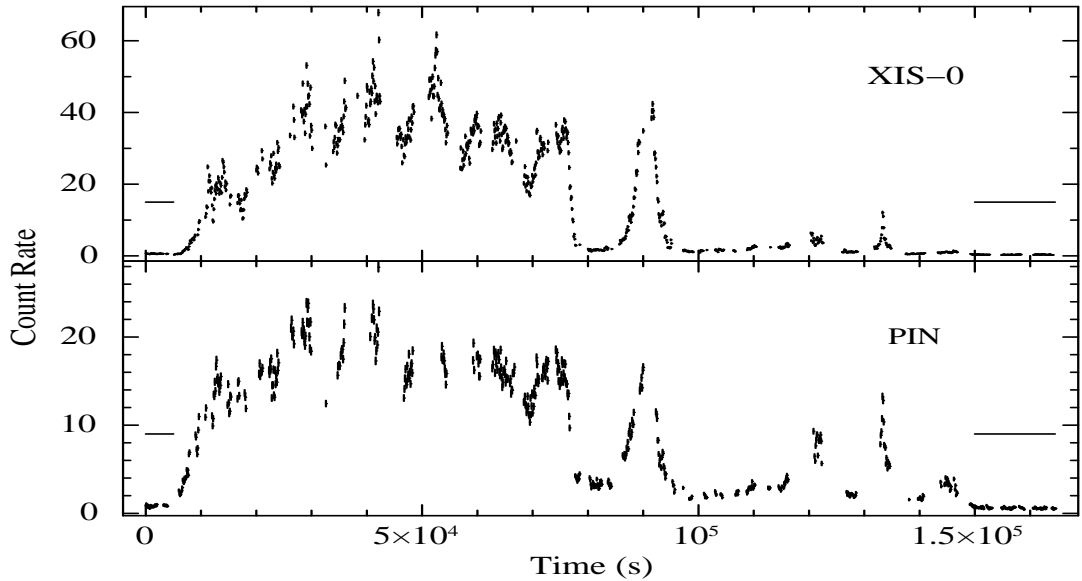}
\caption{Light curves (with 100 s binning time) obtained from the $Suzaku$ 
observation of the high mass X-ray binary pulsar Cen~X-3. Data from XIS-0
and HXD/PIN detectors are plotted here. The observation was carried out 
covering nearly entire orbit of the pulsar. The horizontal lines in the
beginning and end of the XIS and PIN light curves in both the panels show 
the duration of eclipse covered during the $Suzaku$ observation.}
\label{fg1}
\end{figure*}

The broad-band, (0.12--100 keV) out-of-eclipse pulse-phase-averaged spectrum
of Cen~X-3 is generally described by an absorbed power-law plus a broad iron
emission line at $\sim$6.7 keV along with a high energy cut-off at $\sim$14 keV
(Santangelo et al. 1998). A soft excess found in the spectrum below 1 keV has
been interpreted as a black-body with $kT$ $\sim$0.1 keV (Burderi et al. 2000) that 
is now known to be present in many binary X-ray pulsars (Paul et al. 2002;
Naik \& Paul 2004a, 2004b and references therein). A cyclotron resonance 
feature at $\sim$28 keV has also been detected and the corresponding magnetic
field strength close to the surface of the neutron star was calculated to be B 
$\sim$ (2.4--3.0) $\times$ 10$^{12}$ G (Santangelo et al. 1998). A number of
emission lines are expected in the X-ray spectrum of this source due to its 
high luminosity ($\sim$ 10$^{36-37}$ erg s$^{-1}$) and the presence of a strong
stellar wind from the companion star (Ebisawa et al. 1996; Wojdowsky et al. 
2003; Iaria et al. 2005). Investigation of emission lines, in particular
iron lines in 6.4-7.1 keV energy range, at different orbital phases provide
significant information on the physical condition of the system. It is 
understood that the 6.4 keV iron emission line is due to fluorescence of
cold material close to the neutron stars (Nagase 1989) where as the 6.7 keV and 
6.9 keV lines are considered to be originated in the highly photoionized
accretion disk corona (Kallman \& White 1989). ASCA observation of Cen~X-3
clearly resolved three iron emission lines at 6.4 keV, 6.7 keV, and 6.9 keV
during the eclipse and out-of-eclipse data. The line parameters during the
eclipse and out-of-eclipse data confirmed that the 6.4 keV fluorescent line
is emitted from the region close to the neutron star where as the highly
photoionized plasma that emits 6.7 and 6.9 keV lines is more extended than
the size of the companion star (Ebisawa et al. 1996). 

In the present work, we have carried out an extensive spectral analysis 
of the  $Suzaku$ observation of the HMXB pulsar Cen~X-3 covering the eclipse, 
out-of-eclipse and the rarely observed dipping activity. For this purpose, 
we describe the observations, data analysis \& results in the following 
section. Then in the next section, we discuss the results.

\section{Observation, Analysis and Results}

Cen~X-3 was observed with the $Suzaku$ satellite from 2008 December 8 
06:55:36 to 2008 December 10 05:00:19 using the XISs, HXD/PIN and HXD/GSO
detectors. The observation has a total integration time of 97.587 ks. 
$Suzaku$, the fifth Japanese X-ray astronomy satellite (Mitsuda et al. 
2007), covers 0.2--600 keV energy range with the two sets of instruments, 
X-ray Imaging Spectrometers (XIS) covering 0.2-12 keV energy range, and the 
Hard X-ray Detectors (HXD) which covers 10--70 keV with PIN diodes and 
30--600 keV with GSO scintillators. Among the 4 sets of XISs, XIS-1 is 
back illuminated (BI) whereas XIS-0, XIS-2, and XIS-3 are front illuminated 
(FI). The field of view of the XIS is 18'$\times$18' in a full window mode 
with an effective area of 340 cm$^2$ (FI) and 390 cm$^2$ (BI) at 1.5 keV. 
The HXD is a non-imaging instrument that is designed to detect high-energy 
X-rays. The effective areas of PIN and GSO detectors are $\sim$145 cm$^2$ 
at 15 keV and 315 cm$^2$ at 100 keV respectively. For a detailed description 
of the XIS and HXD detectors, refer to Koyama et al. (2007) and Takahashi 
et al. (2007). Cen~X-3 was observed in HXD nominal mode. As XIS-2 is no more 
operational, data from other 3 XISs are used in the present analysis. 

\begin{figure*}
\centering
\includegraphics[height=6.0in, width=7.7in, angle=-90]{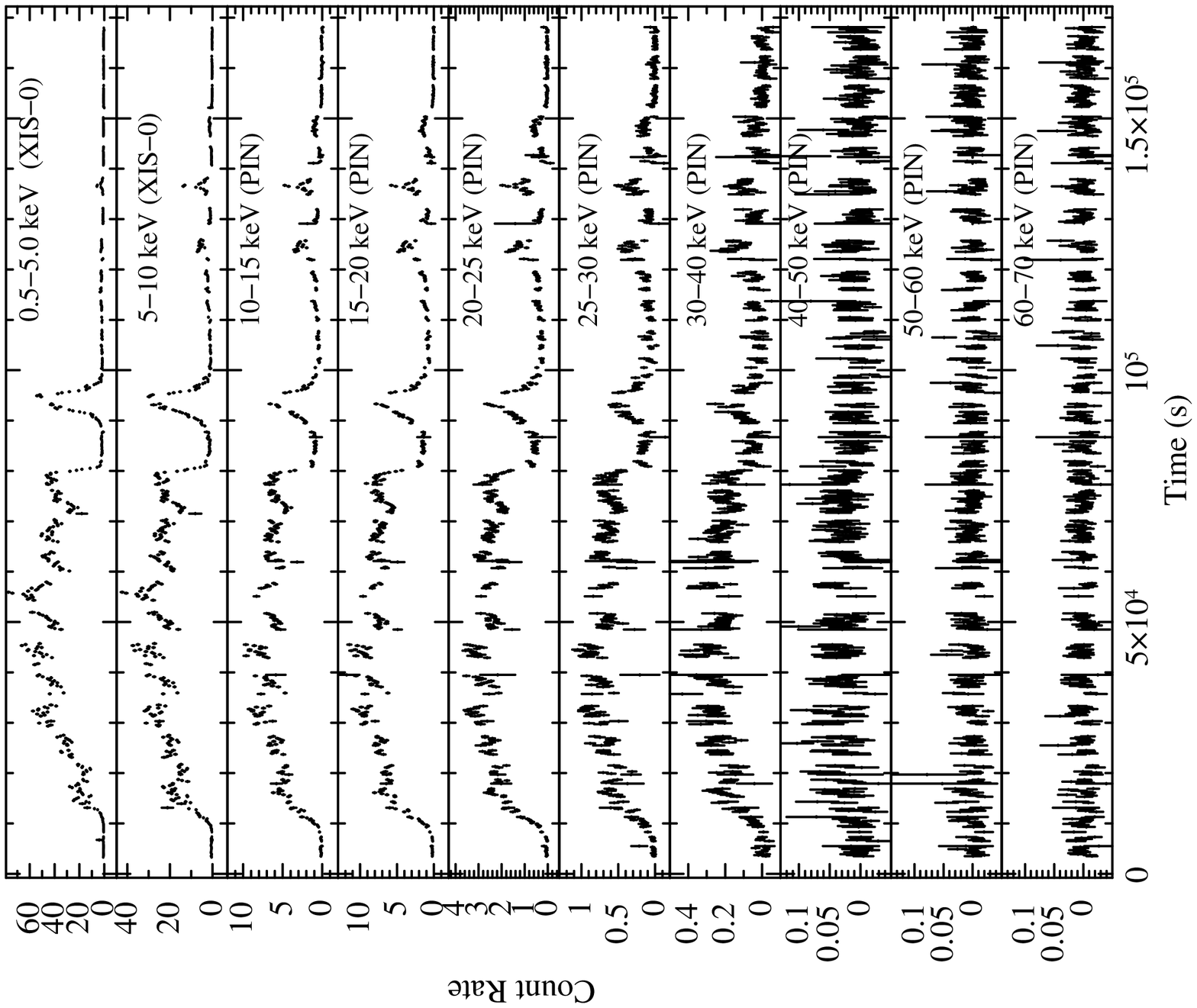}
\caption{Light curve (with 200 s binning time) obtained from the $Suzaku$ 
observation of the high mass X-ray binary pulsar Cen~X-3. Data from XIS-0
and HXD/PIN detectors are plotted here. The observation was carried out 
covering nearly entire orbit of the pulsar.}
\label{erlc}
\end{figure*}

We used public data (ver-2.2.11.22) from a $Suzaku$ observation of the 
pulsar in the present work. We used {\bf heasoft6.5.1} for our analysis.
FTOOLS packages such as $PIN-hxdtime, hxdpi, hxdgrade, XIS-xispi, sisclean$ 
were applied to the unfiltered event files with standard screening criteria 
to create cleaned XIS and PIN event files. The effective exposures, after 
applying the deadtime corrections to the XIS and HXD/PIN data, are found to 
be 48.452 ks and 79.656 ks, respectively. Barycentric correction was then
applied to the reprocessed XIS and PIN event files using $aebarycen$ task 
of FTOOLS. Quick tuned ($bgd_a$) non X-ray background was used for HXD/PIN.
Good time interval (GTI) file was generated for the processed PIN data 
using above background file. For XISs, source region was selected using 
rectangular box regions in such a way so that it covers the entire source. 
The background region for XISs is generated by selecting region
far away from the source. Response file, released in July 16 2008, was 
used for HXD/PIN. Ancillary response file (ARF) and redistribution matrix 
file (RMF) for XISs were generated using $xissimarfgen$ and $xisrmfgen$
of FTOOLs, respectively.

\begin{figure*}
\centering
\includegraphics[height=7in, width=2.3in, angle=-90]{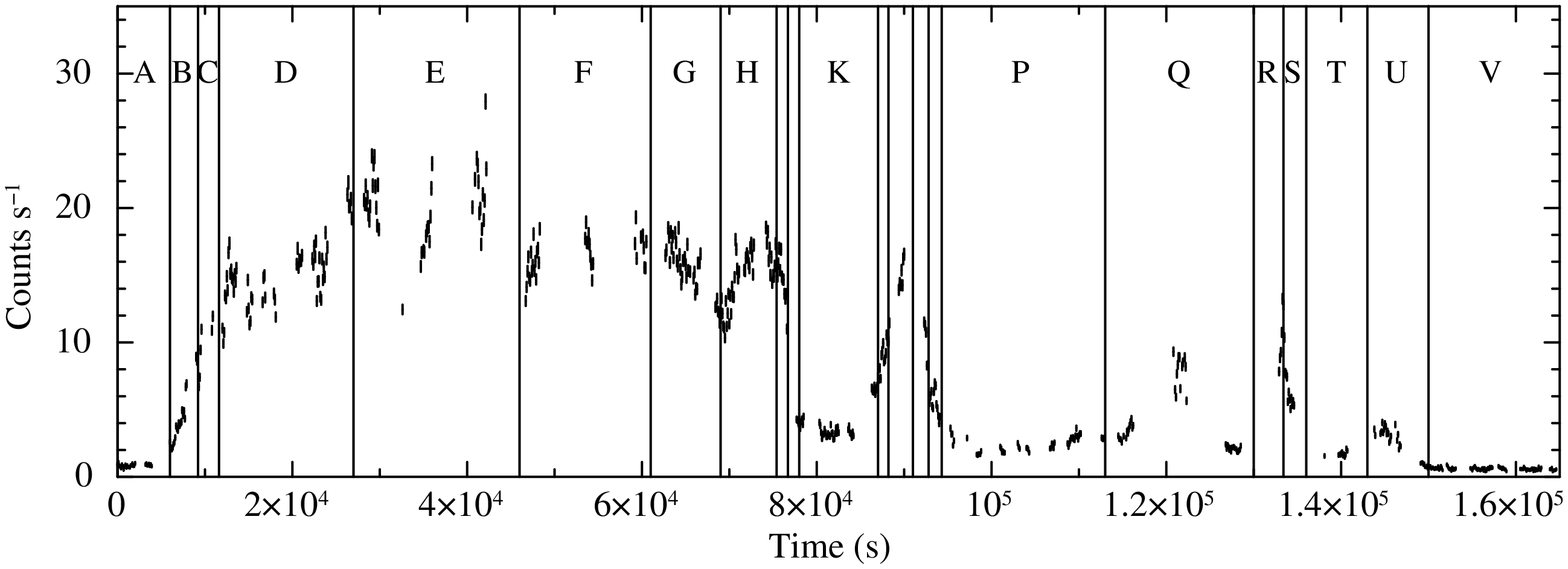}
\caption{HXD/PIN light curve (with 100 s binning time) obtained from the $Suzaku$ 
observation of the high mass X-ray binary pulsar Cen~X-3. The entire light curve
is divided into various segments (marked with alphabets) of different duration
based on the presence of eclipse, dips, and high count rate phase. The smaller
duration regions are not marked with the alphabets in the figure. However, the
nomenclature goes with sequence and the corresponding spectral parameters are
given in Table-1.}
\label{lc_cl}
\end{figure*}

Light curves were extracted from the processed event files for PIN and 
XISs using the standard procedures with 4s and 100 s resolutions. 
Figure~\ref{fg1} shows the XIS-0 and HXD/PIN light curves with 100 s
time resolution. From the figure, it is found that the light curves 
contain distinctive features of eclipsing, out-of-eclipse phase and 
dips. This compelled us to do the time resolved spectroscopy of the 
source in these regions. Source and background spectra were extracted 
from the processed event data for three XISs and PIN. Dead time correction
(using ftool task $hxddtcor$) was applied to the PIN source spectrum. 
To determine the start and end of the eclipse, we folded the light 
curves obtained from the RXTE/ASM dwell data and the $Suzaku$ HXD/PIN 
event data using the orbital period of 2.08706 d (Raichur \& Paul 2008b). 
It is found that the eclipse covers $\sim$0.2 orbital phase range 
($\sim$0.42 days; Raichur \& Paul 2008b) of the binary system. We figured 
out the eclipse in the $Suzaku$ observation to be at the beginning and 
the end of the observation (marked by horizontal lines in both panels 
of Figure~\ref{fg1}). Apart from these two eclipsing parts in the light 
curve, there are several other low X-ray flux regions which are dips. Light 
curves at different energy bands were extracted and after appropriate 
background subtraction are plotted together in Figure~\ref{erlc}. From 
the figure, it can be seen that the dipping feature in the light curve is 
present up to $\sim$40 keV beyond which the light curve seems to be 
feature-less. 

\begin{figure*}
\centering
\includegraphics[height=6.0in, width=3.75in, angle=-90]{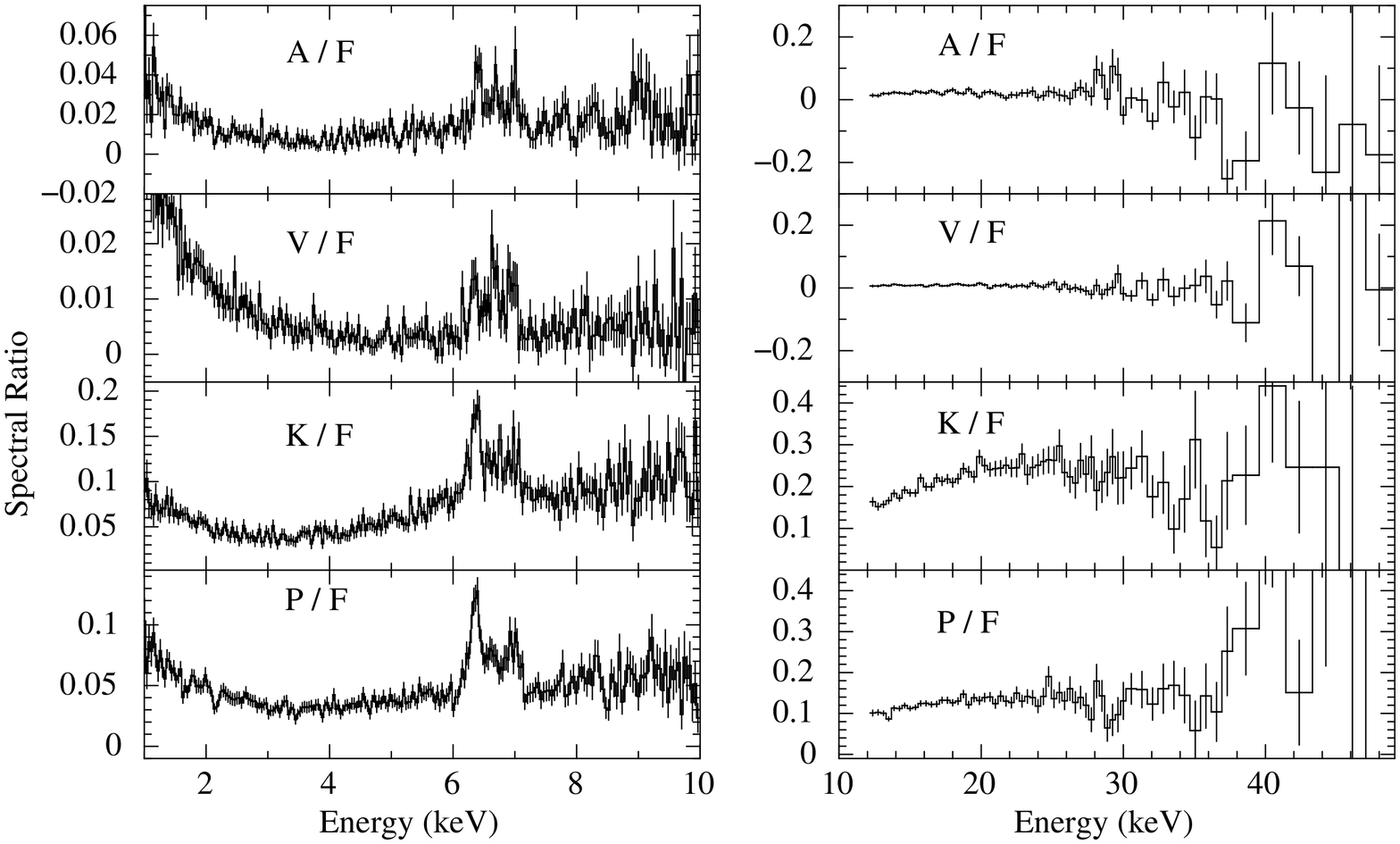}
\caption{Ratios of the background-subtracted spectra of Cen~X-3 during 
the eclipses (marked as 'A' and 'V') and dips (marked as 'K' and 'P') to that 
of the high X-ray intensity segment (marked as 'F') in the light curve (as
shown in Figure~\ref{lc_cl}). The panels in left side are obtained from 
the XIS-0 data where as in right side, obtained from the HXD/PIN data.}
\label{sr}
\end{figure*}

To understand and compare the properties of the X-ray source during the dips, 
eclipse and the rest of the regions, the entire light curve was divided
into different segments viz. eclipse, eclipse-ingress, eclipse-egress,
high count rate region, dip-ingress, dip, dip-egress etc. The light curve 
showing the various segments is plotted in Figure~\ref{lc_cl}. Each alphabet 
in the figure represents certain duration for which the spectral analysis 
was carried out. We extracted XISs and HXD/PIN source spectra for all the 
segments by applying suitable time filters (as shown in Figure~\ref{lc_cl}) 
in $XSELECT$. These spectra contain 2 eclipses, 3 dips, 4 egresses, 5 ingresses,
6 high count regions and 2 bumps with different exposure time and count rates. 
Dead time corrections were applied to all of the 22 HXD/PIN spectra. After 
appropriate background subtraction, simultaneous spectral fitting was done 
using the XIS and PIN spectra with XSPEC V12. All the spectral parameters other 
than the relative normalization, were tied together for all the detectors. Because 
an artificial structure is known to exist in the XIS spectra at around the Si 
edge, we ignored energy bins between 1.75--1.85~keV in the spectral analysis. 
Apart from the Si edge, large fit residuals due to calibration uncertainties 
are often observed near the edge structures of the XIS/XRT instrumental responses.
Therefore, additional model components for possible fluorescence lines at
energies below $\sim$3.5 keV are not considered during the spectral fitting.
The relative instrument normalizations of the three XISs and PIN detectors 
were kept free and the values are found to be in agreement with that at
the time of detector calibration.

In order to evaluate the spectra during the eclipses and dips in the light
curve in a model-independent manner, we normalized them to that of the high
count rate segment (out-of-eclipse and non-dip region). The resulting ratios
in XIS and HXD/PIN energy bands are presented in Figure~\ref{sr}. For simplicity,
we used only XIS-0 data to obtain the spectral ratio and plotted in the figure.
The ratios in XIS energy band indicate the enhancement in the equivalent width
of three iron emission lines during the eclipses and dips in the light curve 
compared to the high count rate segment. Apart from the iron emission lines, 
the ratios also indicate some changes in the shape of the continuum. The ratios 
in HXD/PIN energy band (right panels of Figure~\ref{sr}) show that the photons up 
to $\sim$40 keV are affected by the absorption/obscuration 
during the dips (bottom two panels) in the light curve. During the eclipse 
segments, however, the photons in entire energy band are affected in same manner 
as the ratios (top two panels) are more or less constant. 

To quantify the inference from Figure~\ref{sr}, we have done a simultaneous 
spectral fitting of the XIS and PIN data during the high X-ray count rate 
region to find a suitable model explaining the spectral features in the source.
The broad-band energy spectra (in 0.8-70 keV energy range) were fitted with 
a model consisting a power-law continuum component modified with edge and 
high energy cutoff along with the interstellar absorption, and a Gaussian
function for the iron fluorescence line at 6.4 keV. Presence of line like
features in the residuals at 6.7 keV and 6.9 keV allowed us to add two more
Gaussian functions with the spectral model. Initially we ignored data below 
4 keV. A cyclotron resonance feature at $\sim$30 keV was also found in the 
spectral fitting. However, as the cyclotron features in this pulsar will be 
presented elsewhere, we concentrated on the variation of other spectral 
parameters at different X-ray intensity phases over the binary orbit. Though 
the above model, with the cyclotron resonance feature, fitted well to the high 
count rate spectra, there were inconsistency in the value of other spectral 
parameters during several other segments where the values of high energy 
cut-off and folding energy are found to be $<$4 keV i.e. beyond the fitted 
energy band) with power-law photon-index significantly different to that 
during the high count rate segments. Therefore, the above model was not 
suitable for the spectral fitting to the spectra of all the segments. 

\begin{figure*}
\centering
\includegraphics[height=6.7in, width=2in, angle=-90]{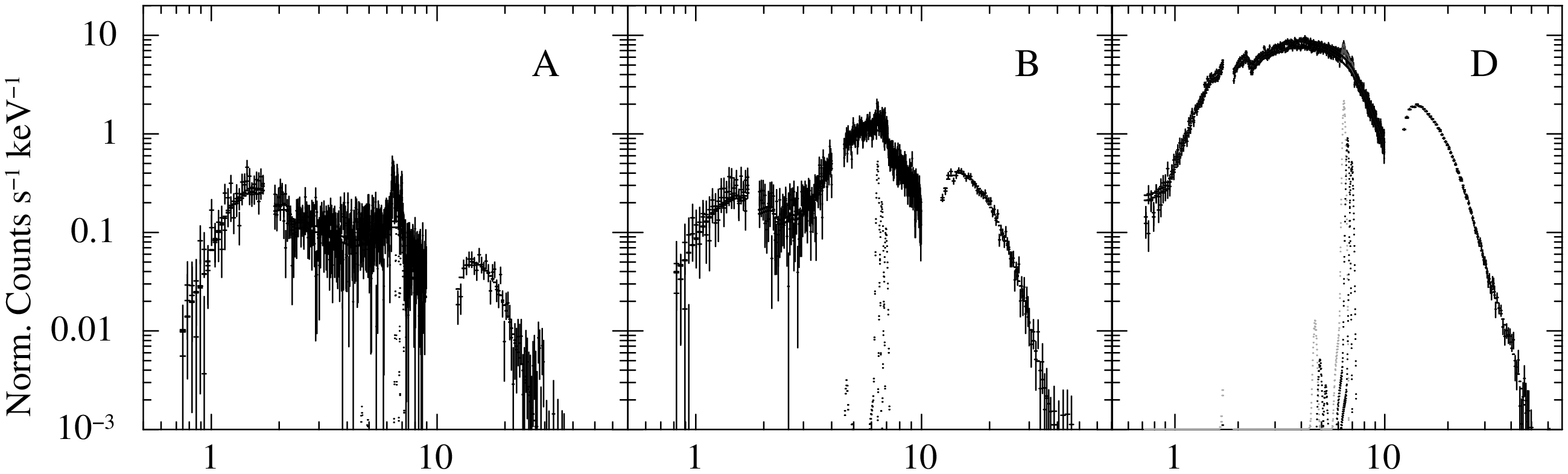}
\includegraphics[height=6.7in, width=2in, angle=-90]{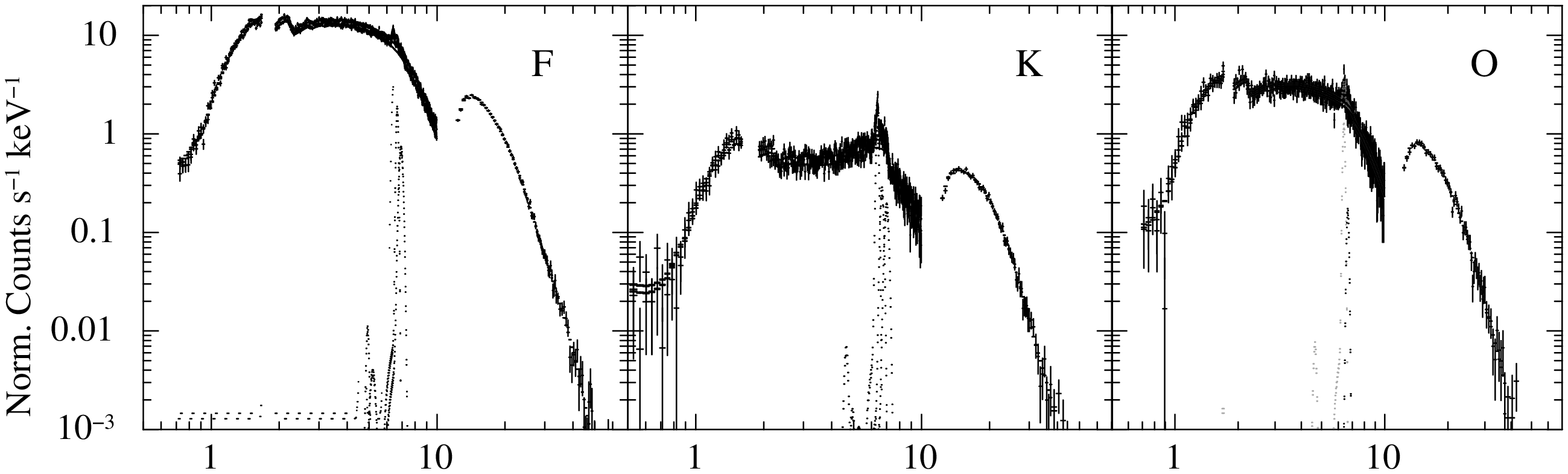}
\includegraphics[height=6.7in, width=2in, angle=-90]{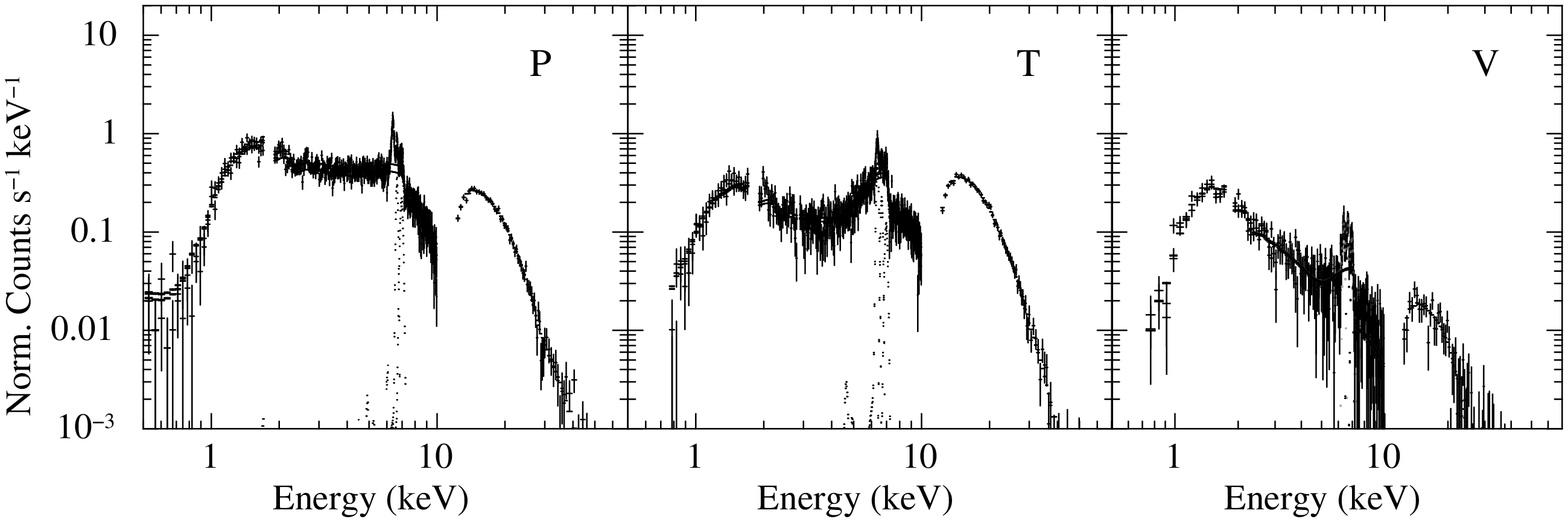}
\caption{Time resolved energy spectra of Cen~X-3 obtained with the XIS-0, 
XIS-3, and PIN detectors of the $Suzaku$ observation, along with the best-fit
model comprising a partially absorbed high energy cutoff power-law continuum
model, three iron line emissions at 6.4 keV, 6.7 keV, and 6.97 keV and cyclotron
resonance features at $\sim$30 keV. The alphabets in the top right corner of
each spectra represent the region from which data were used for spectral fitting
(as shown in Figure~\ref{lc_cl}). The variation in the continuum level and
also the iron emission lines are seen in the spectra.}
\label{spec}
\end{figure*}

We then tried to explore other continuum models to get a better fit to 
the broad band spectrum of Cen~X-3. In the process, we found that a 
partially covering high energy cutoff power-law continuum model (as in
other cases; Naik et al. (2011) and references therein), along with
three Gaussian functions, Galactic absorption, and the cyclotron resonance
feature, fits the pulsar spectrum with acceptable parameters for the spectra 
of all the segments. The partial covering model consists of two power-law 
continua with a common photon index but with different absorbing hydrogen 
column densities. Once we found the suitable continuum model, we extended 
the fitting to low energies ($\sim$0.8 keV) and estimated the spectral 
parameters for all the segments. The time-resolved count rate spectra, 
along with the best-fit model components, of the pulsar Cen~X-3 are
shown in Figure~\ref{spec} for eclipse (A and V), eclipse-egress (B), high 
count rate segments (D and F), dip (K, P and T), dip-ingress (O) segments 
(as shown in Figure~\ref{lc_cl}). In the figure, the iron emission lines
are found to be most prominent during the dips and eclipse segments.

\begin{table*}
\centering
\caption{Best-fit parameters (with 1$\sigma$ errors) of the phase-averaged spectra for Cen~X-3 during the $Suzaku$ observation.}
\begin{tabular}{lllllllllllll}
\hline
\hline
Spec.  &N$_{H1}$  &N$_{H2}$  &Cov.    &Photon	&\multicolumn {2}{|c|}{6.4 KeV line}  &\multicolumn {2}{|c|}{6.7 KeV line} &\multicolumn {2}{|c|}{6.97 KeV line}\\
Reg.    &    &      &Fraction &Index   &Flux$^a$   &Eqw$^b$  &Flux$^a$ &Eqw$^b$  &Flux$^a$  &Eqw$^b$\\
\hline

A	&1.06$\pm$0.03   &88.1$\pm$3.0  &0.94$\pm$0.01  &2.03$\pm$0.02	&0.30$\pm$0.03	&0.43$\pm$0.05	&0.20$\pm$0.02	&0.26$\pm$0.03 	&0.13$\pm$0.03	&0.20$\pm$0.04\\

B	&0.50$\pm$0.04   &32.4$\pm$0.6   &0.96$\pm$0.43	&0.89$\pm$0.05	&0.50$\pm$0.11	&0.08$\pm$.02	&0.24$\pm$0.11	&0.03$\pm$0.02	&0.16$\pm$0.09	&0.03$\pm$0.01\\

C	&2.63$\pm$0.03   &3.9$\pm$0.2    &0.54$\pm$0.01  &0.96$\pm$0.01	&1.80$\pm$0.16	&0.08$\pm$0.01	&0.43$\pm$0.17	&0.02$\pm$0.01	&0$^{+0.16}$	&0$^{+ 0.01}$\\

D	&1.43$\pm$0.01   &2.3$\pm$0.1    &0.64$\pm$0.01  &0.90$\pm$0.01	&2.55$\pm$0.01	&0.08$\pm$0.01	&1.29$\pm$0.09	&0.04$\pm$0.01	&0.90$\pm$0.10	&0.03$\pm$0.01\\

E	&1.01$\pm$0.01   &0.9$\pm$0.1    &0.57$\pm$0.01    &1.03$\pm$0.01	&3.55$\pm$0.10	&0.07$\pm$0.01	&2.69$\pm$0.10	&0.06$\pm$0.01	&1.46$\pm$0.14	&0.03$\pm$0.01\\

F	&1.13$\pm$0.01   &1.1$\pm$0.1     &0.34$\pm$0.01   &1.01$\pm$0.01	&3.78$\pm$0.18	&0.08$\pm$0.01	&1.92$\pm$0.13	&0.04$\pm$0.01	&1.59$\pm$0.13	&0.04$\pm$0.01\\

G	&1.13$\pm$0.01   &0.9$\pm$0.1     &0.29$\pm$0.01 &1.01$\pm$0.01	&3.32$\pm$0.12	&0.09$\pm$0.01	&1.85$\pm$0.13	&0.05$\pm$0.01	&1.41$\pm$0.16	&0.04$\pm$0.01\\

H	&1.09$\pm$0.01   &1.2$\pm$0.1     &0.39$\pm$0.01   &0.98$\pm$0.01	&3.74$\pm$0.14	&0.10$\pm$0.01	&1.68$\pm$0.19	&0.04$\pm$0.01	&0.89$\pm$0.20	&0.02$\pm$0.01\\

I	&1.30$\pm$0.01   &3.8$\pm$0.3     &0.34$\pm$0.01   &1.08$\pm$0.01	&4.12$\pm$0.30	&0.10$\pm$0.01	&0.87$\pm$0.29	&0.02$\pm$0.01	&0.88$\pm$0.30	&0.02$\pm$0.01\\

J	&1.11$\pm$0.02   &3.6$\pm$0.4     &0.32$\pm$0.01   &0.88$\pm$0.01    &2.11$\pm$0.18	&0.16$\pm$0.01	&0.42$\pm$0.13	&0.03$\pm$0.01	&1.15$\pm$0.21	&0.09$\pm$0.02\\

K	&0.95$\pm$0.02   &59.8$\pm$1.0    &0.86$\pm$0.01  &1.20$\pm$0.01	&1.23$\pm$0.05	&0.28$\pm$0.01	&0.43$\pm$0.06	&0.09$\pm$0.01	&0.44$\pm$0.06	&0.10$\pm$0.01\\

L	&1.17$\pm$0.03   &9.7$\pm$0.3     &0.75$\pm$0.01  &0.74$\pm$0.01      &3.63$\pm$0.20	&0.22$\pm$0.01	&1.60$\pm$0.19	&0.09$\pm$0.01	&1.63$\pm$0.22	&0.10$\pm$0.01\\

M	&1.04$\pm$0.01   &1.9$\pm$0.2     &0.28$\pm$0.01  &0.84$\pm$0.01    &5.19$\pm$0.30	&0.17$\pm$0.01	&2.62$\pm$0.29	&0.08$\pm$0.01	&2.92$\pm$0.32	&0.09$\pm$0.01\\

N	&1.18$\pm$0.01   &2.2$\pm$0.2     &0.35$\pm$0.01     &1.10$\pm$0.01	&2.89$\pm$0.26	&0.08$\pm$0.01	&1.28$\pm$0.33	&0.04$\pm$0.01	&0.88$\pm$0.52	&0.02$\pm$0.01\\

O	&1.29$\pm$0.02   &92.9$\pm$5.4     &0.47$\pm$0.01 &1.01$\pm$0.01	&1.30$\pm$0.15	&0.11$\pm$0.01	&0.21$\pm$0.11	&0.02$\pm$0.01	&0.01$^{+0.10}_{-0.01}$	&0.01$\pm$0.01\\

P	&0.92$\pm$0.01   &79.2$\pm$1.9      &0.84$\pm$0.01 &1.32$\pm$0.01	&0.94$\pm$0.04	&0.38$\pm$0.02	&0.32$\pm$0.03	&0.12$\pm$0.01	&0.35$\pm$0.04	&0.14$\pm$0.02\\

Q	&0.80$\pm$0.02   &56.3$\pm$0.7      &0.91$\pm$0.01   &0.92$\pm$0.01	&1.23$\pm$0.05	&0.21$\pm$0.01	&0.57$\pm$0.04	&0.09$\pm$0.01	&0.61$\pm$0.04	&0.10$\pm$0.01\\

S	&0.85$\pm$0.04   &33.9$\pm$0.8   &0.88$\pm$0.01   &0.58$\pm$0.01	&1.61$\pm$0.12	&0.18$\pm$0.01	&1.10$\pm$0.20	&0.12$\pm$0.02	&1.28$\pm$0.16	&0.14$\pm$0.02\\

T	&0.65$\pm$0.02   &85.0$\pm$1.1    &0.96$\pm$0.26   &1.31$\pm$0.01	&0.51$\pm$0.03	&0.24$\pm$0.02	&0.24$\pm$0.04	&0.10$\pm$0.02	&0.20$\pm$0.04	&0.09$\pm$0.02\\

U	&0.66$\pm$0.04   &84.9$\pm$1.9     &0.96$\pm$0.45   &0.83$\pm$0.02	&0.62$\pm$0.04	&0.34$\pm$0.02	&0.28$\pm$0.06	&0.13$\pm$0.03	&0.27$\pm$0.05	&0.13$\pm$0.03\\

V	&1.34$\pm$0.04   &132.2$\pm$4.5   &0.97$\pm$0.64   &2.89$\pm$0.03       &0.10$\pm$0.02	&0.39$\pm$0.06	&0.10$\pm$0.02	&0.35$\pm$0.06	&0.06$\pm$0.01	&0.26$\pm$0.05\\
\hline
V$*$    &1.05$\pm$0.05   &---             &---     &2.38$\pm$0.07       &0.14$\pm$0.01  &0.46$\pm$0.02   &0.14$\pm$0.01       &0.32$\pm$0.01    &0.12$\pm$0.02	  &0.41$\pm$0.02\\
\hline
\hline
\end{tabular}
\begin{flushleft}
N$_{H1}$ = Equivalent hydrogen column density (in 10$^{22}$ units), N$_{H2}$ = Additional hydrogen column density (in 10$^{22}$ units), $^a$ : in 10$^{-11}$  ergs cm$^{-2}$ s$^{-1}$ unit, $^b$ : Equivalent width (in keV).\\ 
$*$ : Best-fit parameters obtained by fitting the eclipse spectrum (segment 'V') by using a high energy cut-off power law model with interstellar absorption and three Gaussian functions for the three iron emission lines.\\
\end{flushleft}
\label{spec_par}
\end{table*}

\begin{figure}
\centering
\includegraphics[height=3.3in, width=3.5in, angle=-90]{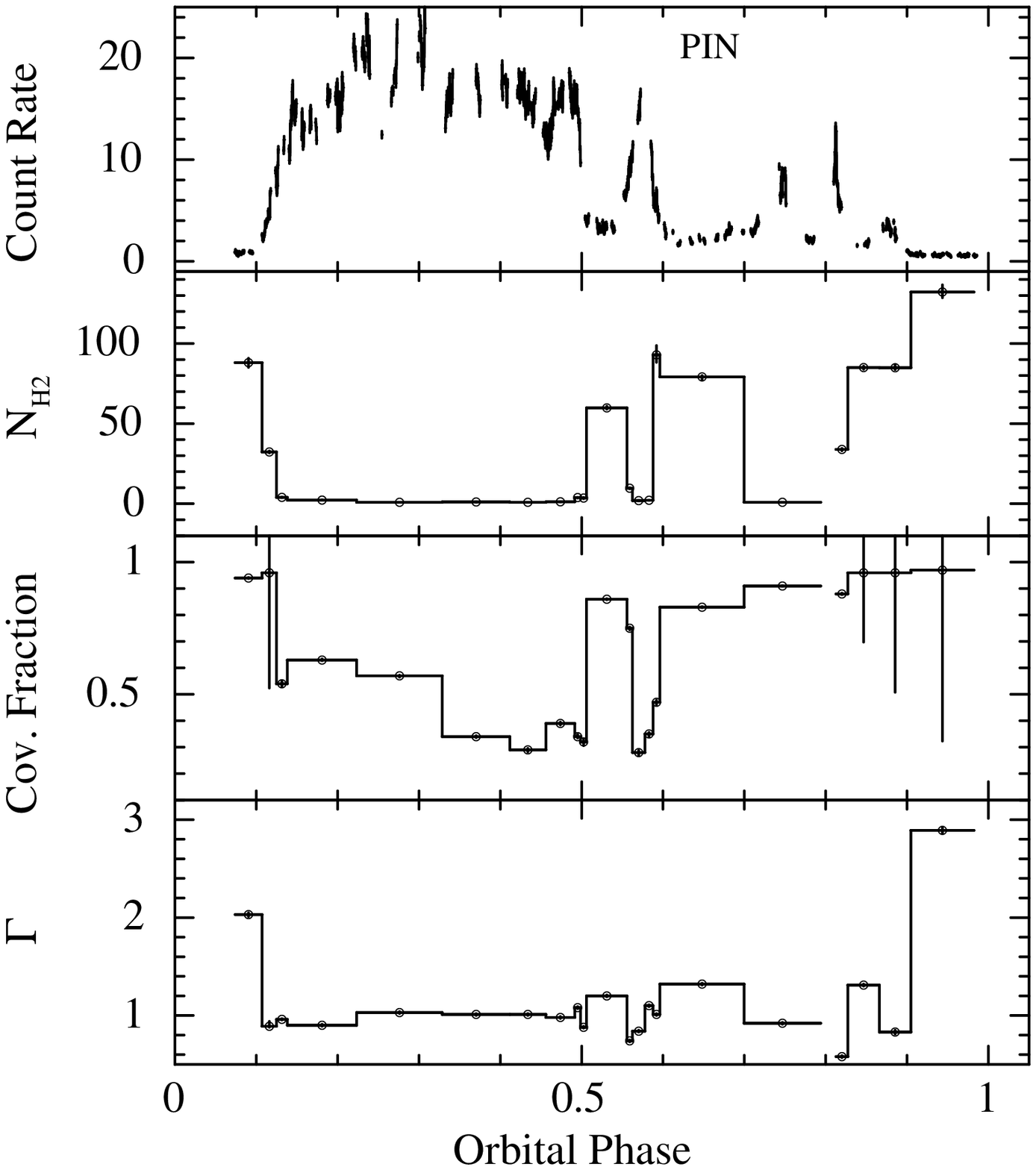}
\caption{The top panel of the figure shows the HXD/PIN light 
curve (with 100 s resolution) of the $Suzaku$ observation of 
Cen~X-3. The mid-eclipse time is considered as orbital phase zero. 
The values of N$_{H2}$ (in 10$^{22}$ units), covering 
fraction, and power-law photon index obtained from the time 
resolved spectroscopy of the $Suzaku$ observation are shown in 
second panel, third panel, and bottom panel, respectively. The 
errors shown in the figure are estimated for 1$\sigma$ confidence 
level.}
\label{trs}
\end{figure}

The best-fit parameters obtained from the simultaneous spectral fitting to 
the XIS and PIN data for all the segments (as given in Figure~\ref{lc_cl}) 
of the $Suzaku$ observation of Cen~X-3 are given in Table~\ref{spec_par}.
The values of the additional absorption column density (N$_{H2}$) are found 
to vary in a wide range starting from $\sim$10$^{22}$ atoms cm$^{-2}$ to 132 
$\times$ 10$^{22}$ atoms cm$^{-2}$. The significantly high values of N$_{H2}$
along with the corresponding covering fractions during dips and the eclipse
segments in the light curve are understood to be because of the 
obscuration/absorption of the X-ray photons from the pulsar by dense matter
along the line of sight. Apart from the significantly high value of N$_{H2}$ 
and covering fractions at the dips and eclipse segments, the iron emission 
line parameters are also observed to be very different from those during the
high count rate segments. The change in values of N$_{H2}$, covering 
fraction and the power-law photon index over the orbital phase of the pulsar
during the $Suzaku$ observation are shown in Figure~\ref{trs}. In the figure, 
the mid-eclipse time (derived from the orbital parameters given in Table~1 of
Paul et al. 2005) is considered as phase zero. From Table~\ref{spec_par} and 
Figure~\ref{trs}, it can be seen that the values of N$_{H2}$ and covering fraction 
for the spectral segment 'O' are different from that of during the dips/eclipse 
segments or high count rate segments. The value of N$_{H2}$ is high and found to 
be comparable to that in the next segment, which is a dip. But the covering 
fraction is only about $\sim$0.5. This is consistent with the scenario of 
progressive covering.

\begin{figure*}
\centering
\includegraphics[height=6.0in, width=4.5in, angle=-90]{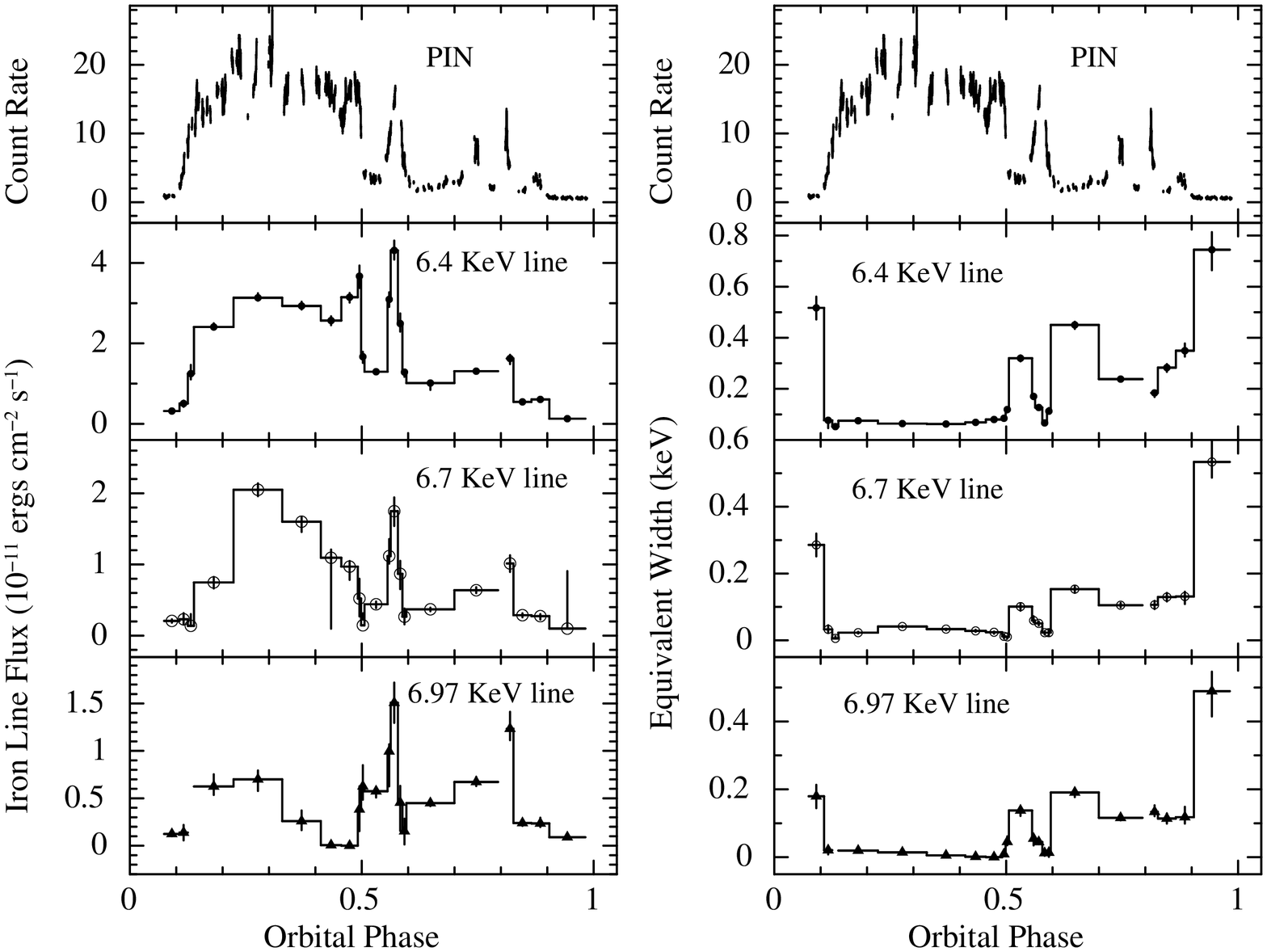}
\caption{Iron emission line parameters obtained from the time resolved
spectroscopy of $Suzaku$ observation of Cen~X-3. The errors shown
in the figure are estimated for 1$\sigma$ confidence level. The top
panels show HXD/PIN light curves with 100 s resolution. The bottom 
three panels in left side show the change in estimated line flux in 
10$^{-11}$ ergs cm$^{-2}$ s$^{-1}$ units where as the bottom three 
panels in right side show the variation in the equivalent widths for 
three iron emission lines during the entire observation.}
\label{trfe}
\end{figure*}

The dependence of the three iron K$_\alpha$ emission line intensities and 
equivalent widths on the X-ray intensity at different segments of the orbit 
is shown in Figure~\ref{trfe}. It is found that the line intensities are 
minimum during the eclipse segments. During the dips, though the lines are 
comparatively more intense than during the eclipse segments, the line flux 
is significantly low compared to the high count rate segments. The equivalent 
width of the 6.4 keV line during dips is found to be comparable to that during 
the eclipse segments which is $\sim$400 eV. However, the equivalent width of 
the 6.7 and 6.97 keV lines during dips were significantly low compared to that 
during the eclipse segments. These results suggest that the iron emission line 
parameters during the dip evolve due to a progressive covering. Compared to the 
6.4 keV line, the 6.7 and 6.97 keV lines evolve slightly differently as they 
originate further away from the neutron star. This indicates that the dipping 
activity during the $Suzaku$ observation of Cen~X-3 is possibly because of the 
presence of structures in the outer region of the accretion disk, as seen in 
low mass X-ray binary systems.

Table~\ref{spec_par} and Figure~\ref{trs} show that the source spectrum
during the eclipse segments ('A' and 'V' in Figure~\ref{lc_cl}) is steep
compared to the rest of the binary phases; the power-law photon index is
as high as 3, where as during the other segments, it is $\sim$1. We attempted
to fit the eclipse spectrum using a high energy cutoff power-law model with
interstellar absorption and three Gaussian functions for the iron emission lines.
The best-fit parameters for the simultaneous spectral fitting to the eclipse
data using this model are given in the last line of Table~\ref{spec_par}. The 
high value of power-law index during the eclipse compared to the rest of the 
orbital phases is seen earlier observations of Cen~X-3 (Thompson \& Rothschild 2009
and references therein). Similar changes in the spectral slope during the
eclipse and rest of the binary phases are also seen in 4U~1538-52 (Robba et
al. 2001). The observed spectral steepness during the eclipse of these objects
are attributed to the interstellar dust scattering. Numerous studies of the
eclipse and out-of-eclipse spectra of these type of objects found that the 
dust scattered component has a spectrum that is steeper by a factor 
proportional to $E^{-2}$ that affects the eclipse spectrum most (Nagase et
al. 2001; Robba et al. 2001; Clark 2004; Thompson \& Rothschild 2009).

\section{Discussion}
The high mass X-ray binary pulsar Cen~X-3, though, has been observed
at many different epochs with various observatories, $Suzaku$ observation 
of the pulsar in 2008 December 8-10 is one of the longest imaging spectroscopic
observation. This observation covers almost the entire orbital period of the 
binary system. Significant flux variation over the orbit has been seen in the 
X-ray light curve of the pulsar. Presence of dips in the light curve which is 
common in the LMXBs and rare in HMXBs, is the key feature of the observation. 
A detailed orbital phase resolved spectroscopy of the pulsar in wide X-ray band 
is, therefore, very important to understand the properties of the binary system. 

Selection of an appropriate continuum model is important to investigate
the presence of several features such as the soft excess represented by 
a blackbody component, emission lines, cyclotron absorption features etc.
in the X-ray binary pulsar spectrum. The phenomenological model which has 
been commonly used to describe the continuum spectra of X-ray binary pulsars
is a combination of a power law of photon index $\sim$1, a black-body 
component with a temperature of few hundreds of eV and a high energy 
cut-off above 10 keV. Over and above, there are emission lines from ions 
and broad absorption lines due to cyclotron resonances (Orlandini 2006). 
In case of a few other X-ray binary pulsars, it has been reported that the 
absorption has two different components (Endo et al. 2000; Mukherjee \& Paul 
2004; Naik et al. 2011 and references therein). In this model, one absorption 
component absorbs the entire spectrum where as the other component absorbs 
the spectrum partially. This model is known as partial covering absorption model. 
Though the power law with high energy cutoff model explains the pulse phase 
averaged pulsar spectra well, it runs into problem while fitting the spectra 
during the dip phases of pulsars showing dips or dip-like features in the 
pulse profiles. However, the partial absorption model fits very well to all 
the pulse phase resolved spectra of these pulsars (Her~X-1 -- Endo et al. 2000,
GRO~J1008-57 -- Naik et al. 2011; 1A~1118-616 -- Devasia et al. 2011a, Maitra 
et al. 2011; GX~304-1 -- Devasia et al. 2011b). This model has also been used to 
describe the orbital phase resolved spectroscopy in some of the high mass X-ray 
binary systems such as Cen~X-3 (present work), GX~301-2 (Mukherjee \& Paul 2004; 
Naik et al. 2009) to explain the variation of the absorption column density 
over the binary orbit. 

In the partial covering model, $N_{\rm H2}$ is interpreted as the column 
density of the material that is local to the X-ray source, while $N_{\rm H1}$ 
accounts for the rest of the material along with the Galactic absorption. 
The covering fraction is defined as $Norm2/(Norm1+Norm2)$ where $Norm1$ and 
$Norm2$ are normalizations of the two power laws respectively. It is seen 
from Table-1 that the $N_{\rm H2}$ and covering fraction remain substantially 
high ($>$ 50$\times$10$^{22}$ units and $>$80\%, respectively) during the dips 
and the eclipse regions in the light curve. This means that there is dense and 
clumpy material present at certain orbital phases of the Cen~X-3 binary system
causing dips in the light curve. A study of iron emission line parameters during
eclipse-ingress, eclipse and eclipse-egress with the $ASCA$ noted that the 6.4 
keV fluorescent line is emitted from a region close to the neutron star where as
the 6.7 keV and 6.97 keV line emission region is much larger and more extended
than the size of the companion star (Ebisawa et al. 1996). This information can
be used to compare the iron line parameters during the dips and eclipse segments
of the $Suzaku$ observation of the pulsar. It is found that the equivalent widths
of the three iron emission lines are significantly larger during the dips and 
eclipse segments in the light curve (Figure~\ref{trfe}). The flux of these lines 
are also considerably low during these orbital phases compared to that of the high 
count rate phases in the light curve. The high equivalent width and low flux of 
iron emission lines at dips and eclipse segments also suggest the presence of 
significant absorption and reprocessing during corresponding orbital phases of 
the binary system. This feature is similar to that seen in Her~X-1 and LMC~X-4 
(Naik \& Paul 2003), two other sources that show strong superorbital X-ray intensity
variations understood to be due to varying degree of obscuration by the precessing
accretion disk. 

During the ASCA observation of Cen~X-3, unlike the present work, the equivalent width 
of 6.4 keV line hardly varied during the eclipse-ingress, eclipse, and eclipse-egress
(Ebisawa et al. 1996). In fact, this is in contrast with the similar HMXBs Vela~X-1
and GX~301-2. During the eclipse, in Vela~X-1, the 6.4 keV iron line equivalent width
increases significantly than that during the out-of-eclipse (Nagase et al. 1994 and
references there in). Though GX~301-2 does not show binary eclipse, a significant 
increase in the 6.4 keV iron emission line equivalent width was observed during an
intensity dip that is thought to be due to the occultaion of the neutron star by
dense matter (Leahy et al. 1988). During the intensity dips in Cen~X-3 (present work), 
it is found that the iron emission line intensity dropped by a factor of 2 to 3 with 
significant increase in the line equivalent widths. However, the drop in continuum 
flux is found to be by an order of magnitude or more during the dips (Figure~\ref{trfe}). 
The decrease in the continuum flux during the dips suggest that the direct power law
component in the partial covering model is almost vanished or completely absorbed
where as the second power law component contributes to the X-ray flux observed during
the dips. Therefore, the presence of dips in the X-ray light curve of Cen~X-3 can
be explained by the eclipse/obscuration of the neutron star by dense matter which
is smaller in size compared to the scattering region surrounding the neutron star,
as seen in GX~301-2.

\section*{Acknowledgments}
The research work at Physical Research Laboratory is funded by the Department 
of Space, Government of India. We express our thanks to the anonymous referee
for his/her valuable comments which improved our paper. The authors would like 
to thank all the members of the $Suzaku$ for their contributions in the instrument
preparation, spacecraft operation, software development, and in-orbit 
instrumental calibration. This research has made use of data obtained 
through HEASARC Online Service, provided by the NASA/GSFC, in support 
of NASA High Energy Astrophysics Programs.

\end{document}